\documentclass[onecolumn,pra,12pt]{revtex4-1}

\usepackage{amsmath}  
\usepackage{graphicx} 
\usepackage{caption}
\usepackage{subcaption}
\usepackage{dcolumn}
\usepackage{bm}

\begin{document}
\centerline{\LARGE{\textbf{Relativistic Quantum Backflow}}}
\vskip 5mm
\centerline{\large {J. Ashfaque $^1$, J. Lynch$^2$ and P Strange$^2$}}
\vskip 5mm
\centerline{$^1$Max-Planck Institute for Software Systems, Campus E1 5, 66123 Saarbr\"ucken, Germany.}
\centerline{$^2$School of Physical Sciences, University of Kent,} \centerline{Canterbury, Kent, CT2 7NH, UK.}
\vskip 5mm
\centerline{\textbf{Abstract}}
In this paper we discuss relativistic quantum backflow. The general theory of relativistic backflow is written down and it is shown that the backflow can be written as a function of a simple parameter $\epsilon$ which is defined in terms of fundamental constants and the backflow period. Backflow eigenfunctions are determined numerically for a range of values of $\epsilon$ and an explicit expression for the relativistic backflow eigenvalue in terms of the non-relativistic backflow constant is presented. Then backflow eigenvectors are fitted with some standard functions which lead to substantially higher backflow than has been found previously with fitting procedures, for some values of $\epsilon$. In analysing the non-relativistic limit of the theory we show that this problem is one of those rare cases where the relativistic theory is intrinsically more simple than the non-relativistic theory.

\vskip 10mm
\section{\label{Intro}Introduction}
Quantum back flow is a remarkable and yet relatively unknown phenomenon that occurs in quantum mechanics. It is the initially puzzling effect that for a free particle described by a wavefunction, localised in $x<0$ and containing only positive momenta, the probability of remaining in $x<0$ can actually increase with time. Probability can flow backwards, i.e. in the opposite direction to the momentum in certain cases. Allcock \cite{allcock} was the first to identify backflow in his study of the arrival time problem in quantum mechanics. However it was Bracken and Melloy \cite{bracken} who first studied non-relativistic quantum backflow systematically and showed that, although a period of backflow can be arbitrarily long, the increase in probability for remaining in $x<0$ cannot exceed a limited amount given by a dimensionless number which they calculated to be approximately $c_{bf}=0.04$. This value has been refined and the most precise estimate now was computed numerically by Penz and co-workers \cite{penz} to be $c_{bf}=0.0384517$.  A number of authors have provided further insight into this topic.  Aharonov \cite{yuri} and Muga \cite{muga1, muga2} and co-workers  discussed in some detail the apparent contradictions between theory and practice in arrival time measurements. More recently Berry \cite{mvb} has discussed backflow probability and the structure of the boundaries between forward and backflowing regions and their time evolution. He also showed that special wavefunctions with relatively large backflow could be found, but that these could not be sustained for long times. While most investigations of quantum backflow previously have considered free particles, Bostelmann and co-workers have recently extended the theory to systems in which there is a scattering potential \cite{bost}. Halliwell \cite{hall2} has considered a novel approach to the arrival time problem and shown that backflow provides an interesting example in which to test their formalism. They have also discussed backflow in terms of the Leggett-Garg inequalities.  

Melloy and Bracken later generalised their work to the relativistic case \cite{melloy} where the theory is fundamentally different. They solved the backflow eigenvalue equation numerically and showed that the maximum backflow $c_{rbf}$ depends on a dimensionless constant $\epsilon$ which depends on the mass and period of backflow as well as the speed of light and Planck's constant.  They were able to obtain good numerical values for the backflow eigenvalues and eigenvectors as a function of $\epsilon$. The non-relativistic limit is taken as the speed of light $c\rightarrow \infty$ which means  $\epsilon \rightarrow 0$ and in this limit the relativistic backflow eigenvalue $c_{rbf} \rightarrow c_{bf}$. Relativistic backflow has also been investigated by Su and Chen \cite{such}. Their very interesting paper looks at how relativistic backflow can be interpreted in terms of pilot wave theory.

A number of authors have attempted to find analytic expressions for the maximum backflow eigenvector by fitting to the numerical values.  This has proved difficult, but some progress has been made. Yearsley {\it et al} have provided expressions based on Fresnel integrals which yield up to about 70\% of the maximum possible backflow. O'Mullane \cite{om} has attempted to fit Bessel functions to the numerical eigenvector and found an expression that yielded of order 20\% of the maximum, but which suffered from numerical instabilities which made results with higher backflow unreliable. 

\section {Theory}

In this section we write down the standard relativistic quantum theory underlying the backflow problem and then follow the work of Melloy and Bracken \cite{melloy}. We consider a Dirac particle in one dimension. In this case the Dirac wavefunction is a two component quantity and the Dirac equation itself may be written
\begin{equation}
i\hbar \frac{ \partial \psi (x,t)}{\partial t} = {\hat H} \psi(x,t)
\label{dir1}
\end{equation} 
with 
\begin{equation}
\psi (x,t) = \left(
\begin{matrix}
\psi_1(x,t) \cr
\psi_2(x,t) \cr
\end{matrix}
\right)
\end{equation}
and
\begin{equation}
{\hat H}=c\sigma_1 {\hat p} +\sigma_3 mc^2
\label{dirac}
\end{equation}
where $\sigma_j$ are the usual Pauli spin matrices and other symbols have their conventional meaning \cite{ps1}. Of course, in relativistic quantum theory time and space should be treated on an equal footing. In this paper we have written the theory such that the backflow occurs in the spatial coordinates. The theory can be written in a way that allows backflow to occur in the time coordinate. This has been investigated by Su and Chen \cite{such} and they have shown that the theory then necessarily involves negative energy states which can be regarded as positive energy states moving backwards in time. The probability density is
\begin{equation}
\rho(x,t) = \psi^{\dagger} (x,t) \psi(x,t)=\psi_1^*(x,t)\psi_1(x,t)+\psi_2^*(x,t)\psi_2(x,t)
\end{equation}
and the current density is given by
\begin{equation}
j(x,t)=c\psi^{\dagger} (x,t) \sigma_1\psi(x,t)=c(\psi_1^*(x,t)\psi_2(x,t)+\psi_2^*(x,t)\psi_1(x,t))
\label{curr}
\end{equation}
Positive energy solutions to equation (\ref{dir1}), normalised to a distance $L$, are 
\begin{equation}
\psi (x,t)=\frac{1}{\sqrt{L}}\left(
\begin{matrix}
U_1 (p)\cr
U_2 (p)\cr
\end{matrix}
\right)
\exp \left(i(px-E(p)t)/\hbar \right)
\label{eigen}
\end{equation}
where
\begin{equation}
E(p)=+\sqrt{p^2c^2+m^2c^4},\hskip 5mm U_1(p)=\sqrt{\frac{\gamma (p)+1}{2\gamma (p)}},\hskip 5mm U_2(p)= 
\sqrt{\frac{\gamma (p)-1}{2\gamma(p)}}
\end{equation}
and $\gamma (p)=(1-v^2/c^2)^{-1/2}$ is the usual relativistic factor. A complete theory would allow positive and negative energy states in the initial state, but for simplicity and to maintain backflow as a spatial effect only we have restricted ourselves to positive energy solutions. It should also be noted that equation (\ref{eigen}) is not an eigenfunction of the $S_z$ operator so its spin cannot be defined.  Next we set up a wavepacket composed of these solutions
\begin{equation}
\Psi (x,t)=\frac{1}{\sqrt{2\pi\hbar}}\int_0^{\infty}\Phi (p) \exp \left(i(px-E(p)t)/\hbar \right) dp
\label{wfun1}
\end{equation}
The limits on the integral here restrict us to positive momentum only. Here
\begin{equation}
\Phi (p)=f(p)\left(
\begin{matrix}
U_1 (p)\cr
U_2(p) \cr
\end{matrix}
\right) =f(p)U(p)
\end{equation}
and it is easy to show that normalisation of $\Psi(x,t)$ requires
\begin{equation}
\int_0^{\infty}\Phi^{\dagger} (p) \Phi (p) dp=\int_0^{\infty}f^*(p)f(p)dp=1
\end{equation}
$f(p)$ is an envelope function that can be chosen arbitrarily. The next step is to substitute the wavefunction (\ref{wfun1}) into the expression for the current. This results in the following expression for the current density at $x=0$. 
\begin{equation}
j(0,t)=\frac{mc^2}{4 \pi\hbar}\int_0^{\infty}(U_1(p) f^*(p)e^{i\gamma (p) mc^2 t/\hbar}+U_2(p) f(p)e^{-i\gamma (p) mc^2 t/\hbar} +c.c. ) dr
\label{curr1}
\end{equation}
The full theory of relativistic backflow has been written down by Melloy and Bracken \cite{melloy} and we only quote the results here. It turns out to be convenient to write the results in terms of the following dimensionless parameters (Both $p$ and $q$ are momenta)
\begin{eqnarray}
p & =\sqrt{\frac{4m\hbar}{T}}r, \hskip 10mm q =\sqrt{\frac{4m\hbar}{T}}s, \hskip 10mm \epsilon =\sqrt{\frac{4\hbar}{mc^2T}},\cr E(p) & =\sqrt{p^2c^2+m^2c^4}=mc^2\sqrt{1+\epsilon^2r^2}=\gamma (r)mc^2
\label{param}
\end{eqnarray}
where $T$ is the time period over which we measure the current density. The second expression here exhibits the relationship between the $\epsilon$ parameter and the relativistic $\gamma$-factor. The problem now is to choose $f(p)$ such that the current in the negative $x$-direction is maximised subject to the normalisation. This is a straightforward variational principle calculation and results in 
\begin{equation}
\int_0^{\infty}K(r,s)\eta (s)ds=\lambda \eta (r)
\label{eigen1}
\end{equation}
where $\lambda$ is a Lagrange multiplier, the kernel is
\begin{equation}
K(r,s)=-\frac{1}{\pi}\frac{r(\gamma (s)+1)+s(\gamma(r)+1)}{\sqrt{\gamma (r)(\gamma (r)+1)\gamma (s)(\gamma (s)+1)}}
\frac {\sin ({2(\gamma (r)-\gamma (s))/\epsilon^2)}}{(2(\gamma (r)-\gamma (s))/\epsilon^2)}
\label{kern}
\end{equation}
and
\begin{equation}
\eta (r)=e^{-2i\gamma (r)/ \epsilon^2}f(mc\epsilon r)
\label{eta}
\end{equation}
Equation (\ref{eigen1}) is an eigenvalue equation which we must solve to find $f(mc\epsilon r)$. If we let $\epsilon \rightarrow 0$ this is equivalent to letting $c\rightarrow \infty$ and we get the non-relativistic equation 
\begin{equation}
\frac{1}{\pi}\int_0^{\infty}\frac{\sin(r^2-s^2)}{r-s}\eta (s)ds=\lambda \eta (r)
\label{nr}
\end{equation}
for which we know the highest magnitude negative eigenvalue is -0.0384517. If we define the flux
\begin{equation}
\Delta=\int_0^Tj(0,t)dt
\label{flux}
\end{equation}
it follows from the Euler-Lagrange equation that $\Delta=\lambda$ \cite{melloy}. The problem now is how to solve equation (\ref{eigen1}). There is no obvious analytic solution and an analytic solution to the non-relativistic equation ({\ref{nr}) has also remained elusive. Yearsley, Halliwell and co-workers \cite{year1,year2,hall1} have tried a number of approximate solutions to equation (\ref{nr}) which have provided wavefunctions with modest backflow. 

Equation (\ref{eigen1}) can be solved numerically of course. The most efficient method for the non-relativistic case was developed by Penz {\it et. al.} \cite{penz} and is based on the power method for finding maximal eigenvalues and eigenvectors. We have used this method for the relativistic case.

\section{Results}

\subsection{The Eigenvalue}

It turns out that the maximum allowed backflow depends only on $\epsilon$. Although $T$ does not appear in equation (\ref{curr1}) it is determined by the parameters in equations (\ref{param}). As $c\rightarrow \infty$, $\epsilon \rightarrow 0$ and we tend to the non-relativistic limit and indeed the maximum permitted backflow does approach its non-relativistic value. Interestingly, letting $\hbar \rightarrow 0$ also means $\epsilon \rightarrow 0$ and this also yields the non-relativistic value of backflow. Backflow is definitely a quantum effect and we might have expected it to tend to zero as $\hbar \rightarrow 0$ but this is not the case. In fact taking $\hbar \rightarrow 0$ is too simple a way to take the classical limit here. 

We have calculated the maximum backflow as a function of $\epsilon$. The calculation details are as follows. To evaluate equation (\ref{eigen1}) we choose a maximum value of $s$ which we call $q_0$ and a number of points $N_0$ in the region $0\leq s\leq q_0$ and evaluate the integral numerically. We then choose integers $h=1,2,3,\cdots$ and multiply the upper limit of the integral by $\sqrt{h}$ and the number of points at which we evaluate the integrand by $h$. This both increases the range of integration and the density of points in the range simultaneously. The integrals are performed using the matlab trapz function. We increase $h$ progressively until the integral has converged satisfactorily. With a judicious choice of initial values the convergence can be fairly rapid. Currently we are limited to four-figure accuracy in the determination of the maximum (negative) eigenvalue of the backflow operator. Our results are shown in table 1 for some particular values and in graphical form in Figure 1. These values are within 0.001 of the values originally determined by Melloy and Bracken \cite{melloy} where they can be compared. At the lower values of epsilon we are limited by the increasing difficulty of convergence of the backflow eigenfunction.
\begin{table}[htp]
\begin{center}
\begin{tabular}{|c|c|c|c|c|c|c|c|}
\hline
$\epsilon$ & $0.10$ & $0.50$ & $0.80$ & $1.00$ & $1.60$ & $2.00$ & $2.50$ \cr
\hline
$c_{rbf}$ & $0.03686$ & $0.03088$ & $0.02722$ & $0.02498$ & $0.01947$ & $0.01660$ & $0.01372$\cr
\hline
\end{tabular}
\caption{Table showing the highest magnitude negative eigenvalue of the backflow operator which is equivalent to the flux of equation (\ref{flux}). The numerical uncertainty on the numbers is $\pm 2$ in the last digit.}
\end{center}
\label{tab1}
\end{table}
\begin{figure}[h]
\centering
\includegraphics[width=0.90\textwidth]{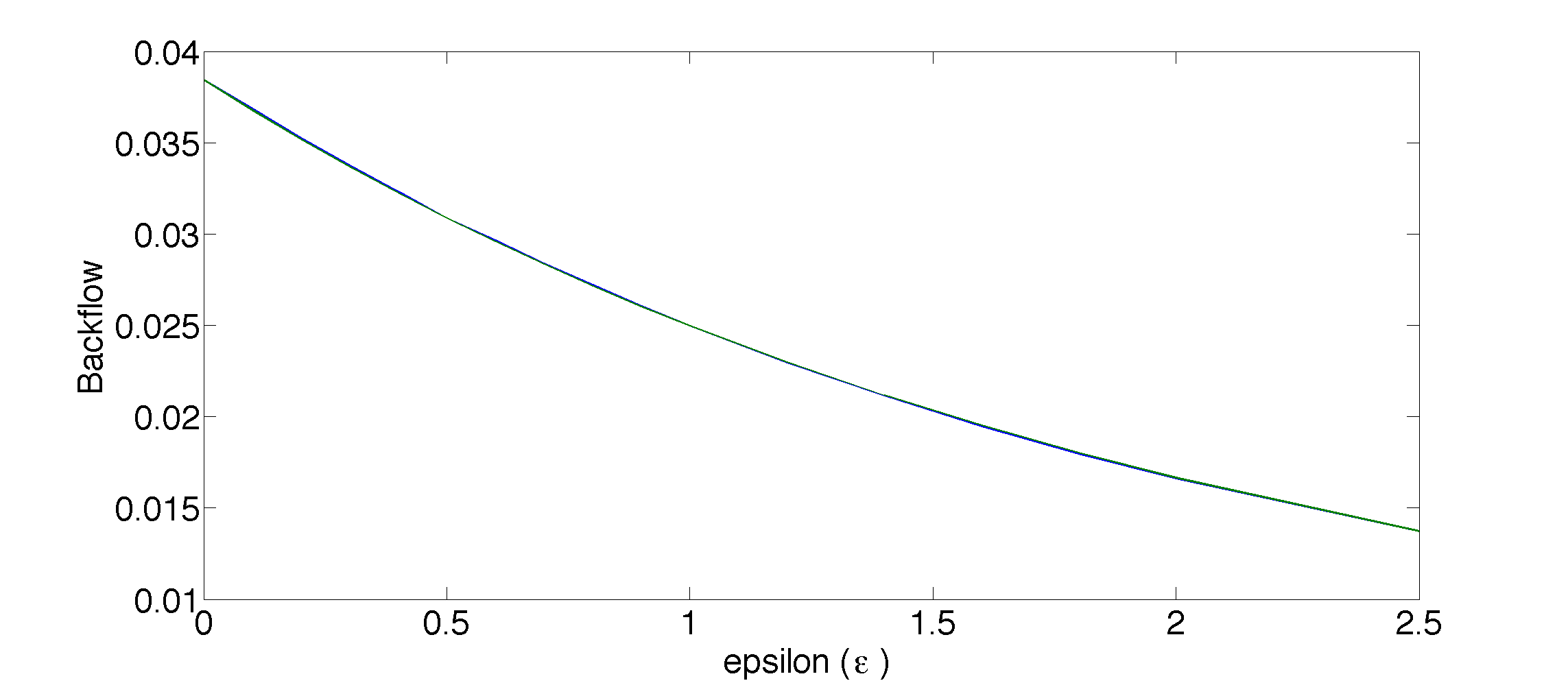}
\caption{There are two lines on this graph: (i) the highest magnitude negative eigenvalue of the backflow operator, which is equivalent to the flux, as a function of $\epsilon$. (ii) The fit to (i) given by equation (\ref{fit}).}
\label{fig1}
\end{figure}
The maximum negative flux as a function of $\epsilon$ can be fit by
\begin{equation}
\Delta =c_{rbf}= c_{bf} \exp \left[ -\frac{4\epsilon}{9}\left(1-4\alpha \epsilon\right)\right]
\label{fit}
\end{equation}
where $\alpha$ is the fine structure constant. This function is also shown in Figure 1.

\subsection{The Eigenvectors}

The eigenvectors $f(r)$ are found from equations (\ref{eigen1}) and (\ref{eta}) numerically and vary with $\epsilon$. We have calculated them and two examples are shown in Figure \ref{fig2}. On the left is the eigenvector evaluated for $\epsilon =0.2$ which is tending towards the non-relativistic limit and on the right is the eigenvector for $\epsilon=1.0$ in which relativity plays a more significant role. Note the differing horizontal scales. It is clear that the frequency of the oscillations becomes substantially more rapid in the non-relativistic limit.

\begin{figure}[h]
\centering
\includegraphics[width=0.45\textwidth]{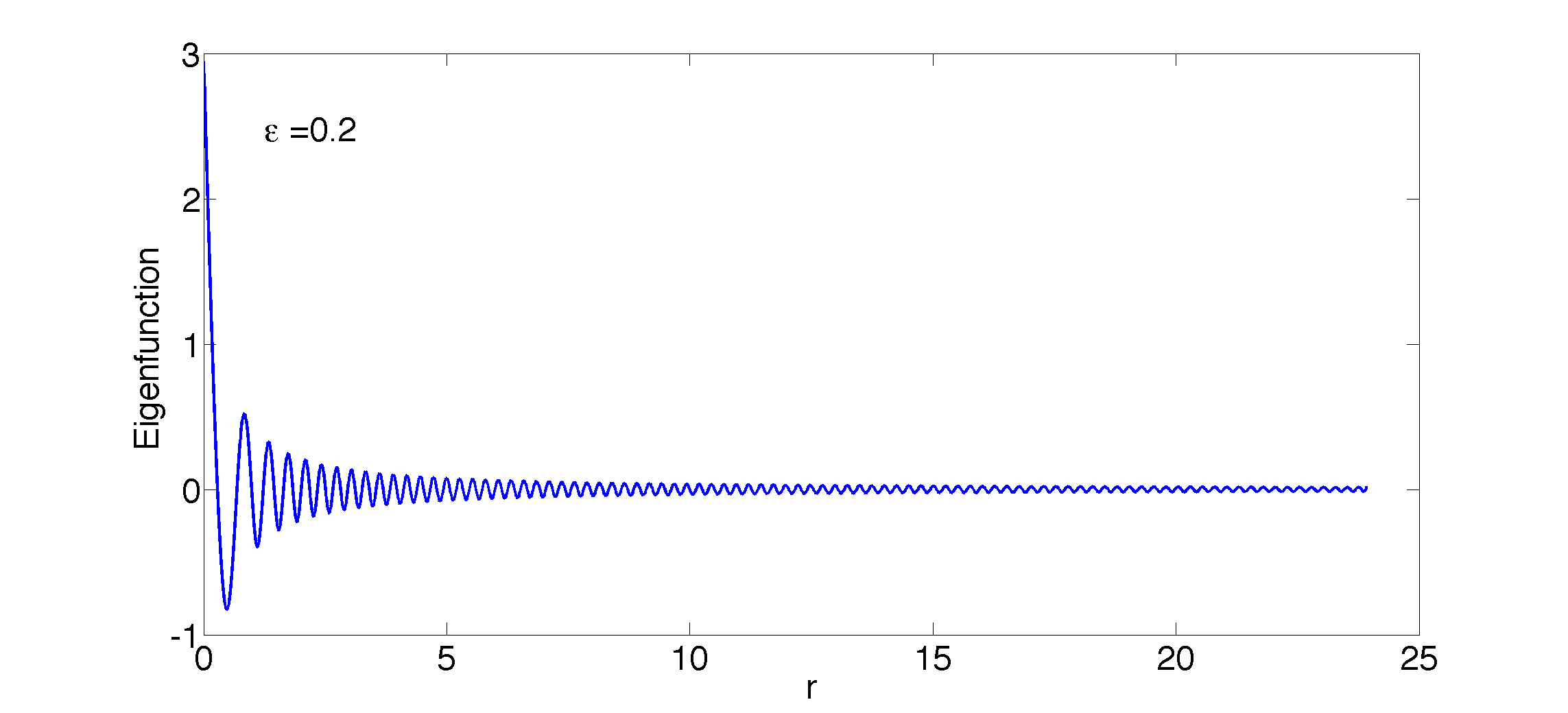}\includegraphics[width=0.45\textwidth]{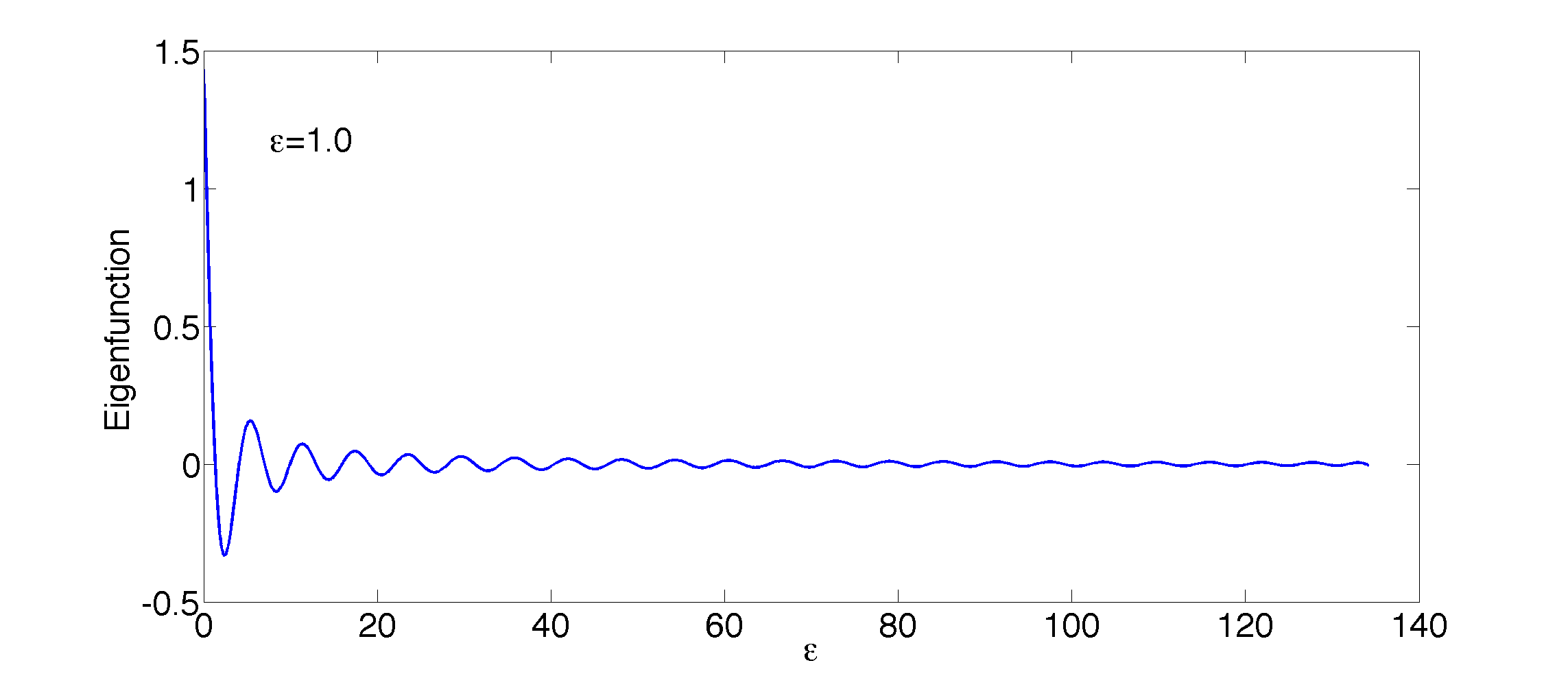}
\caption{The eigenfunction for $\epsilon=0.2$ and $\epsilon=1.0$ determined numerically from equation (\ref{eigen1}).}
\label{fig2}
\end{figure}

In the following subsections we discuss how to find eigenvectors that maximise the backflow. We have done this in two ways. Firstly we have set up a trial function and varied parameters therein to maximise the backflow. Secondly we have set up a trial function and varied the parameters therein to get the best fit to the numerically determined eigenvector.  These two are different and may well not coincide. This has been done for a number of well-known functions. Here we only present the results for Airy functions and Bessel functions of the first kind. In both cases below the parameters $a_j$ were restricted for computational reasons to $-10\leq a_1\leq 0$ and to $0\leq a_j \leq 10$ for all other $j$. 

\subsubsection{Airy Functions}
We have chosen a function of the form
\begin{equation}
f_j(r)=\frac{Ai(x)}{(a_4r+a_5)^{a_6}} 
\label{afit}
\end{equation}
with 
\begin{equation}
x=a_1(r+a_2)^{a_3}
\end{equation}
This form was chosen arbitrarily simply because the numerically determined eigenfunctions do bear a superficial resemblance to the Airy function. Our procedure is to vary all the $a_j$ to find the maximum backflow and the best fit to the eigenfunction. We do the latter by choosing the $a_j$ randomly and then using a curve fitting routine to optimise them to get the best fit to the numerically determined eigenfunction. In general this was done 5000 times for each value of $\epsilon$ but it was found that considerably fewer samplings were required for high values of $\epsilon$. For large values of $\epsilon$ the fit is easy with one well-defined minimum in the parameter space. As we decrease $\epsilon$ the parameter space becomes more complex. At $\epsilon \approx 0.7$ a number of minima of approximately equal depth emerge and it becomes very difficult to be certain we have found the global minimum.  
\begin{figure}[h]
\centering
\includegraphics[width=0.90\textwidth]{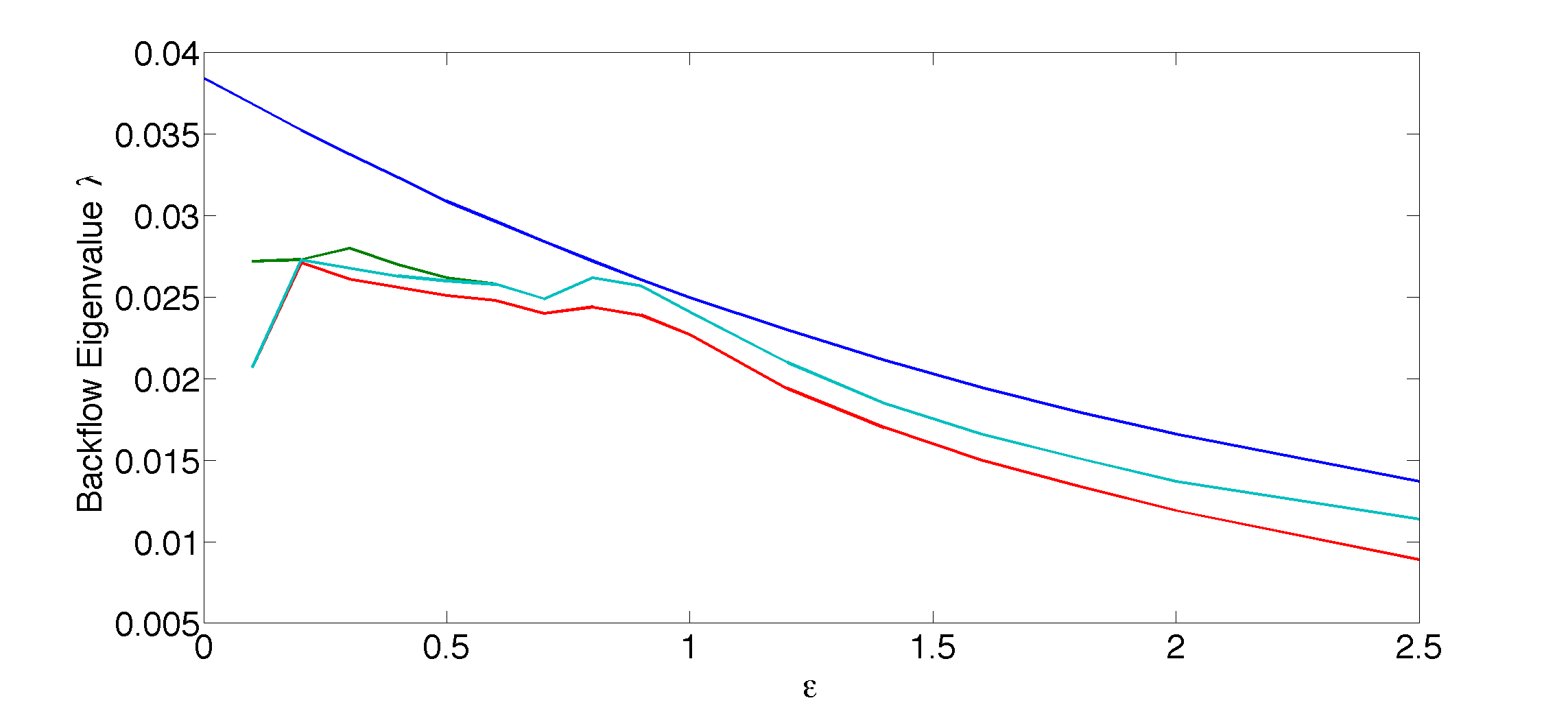}
\caption{{\it Upper line (dark blue)}: the highest magnitude backflow eigenvalue as a function of $\epsilon$ calculated using the exact eigenvector, {\it second line (green)}: The highest magnitude backflow eigenvalue calculated using an optimised value of  equation (\ref{afit}) as the eigenvector, {\it third line (cyan)} the backflow given by the best fit of the form of equation (\ref{afit}) to the numerically determined eigenfunction, {\it fourth line (red)} the backflow given by the best fit of the form of equation (\ref{afit}) to the numerically determined eigenfunction with $a_6=2/3$ }
\label{fig3}
\end{figure}
The results for the eigenvalue are shown in Figure 3. The upper blue line is the theoretical maximum backflow determined numerically.  The green line is the maximum backflow attained with an eigenfunction of the form of equation (\ref{afit}). The cyan line is the backflow eigenvalue calculated using the best fit of the form of equation (\ref{afit}) to the numerically determined eigenfunction. In the relativistic regime above $\epsilon \approx 0.6$ the green and cyan lines coincide to within the level of the numerical accuracy of the calculation. Below $\epsilon \approx 0.6$ it is possible to find a function of the form of equation (\ref{afit}) that gives greater backflow than the best fit to the numerical eigenfunction and so the green and cyan lines diverge for low values of $\epsilon$. The red line is the same as the cyan line except that we have set $a_6=2/3$. For all values $\epsilon >0.2$ this tracks the cyan line, but is a little below it as one would expect. Surprisingly the green, cyan and red lines come together at $\epsilon=0.2$. 

\subsubsection{Bessel Functions of the First Kind}

We have tried fitting the numerically determined eigenfunctions with a number of special functions. The best results are with a simple Bessel function of the first kind with quantum number $l=0$. We have chosen a function of the form
\begin{equation}
f_j(r)=\frac{J_0(x)}{(a_4r+a_5)^{a6}} 
\label{bfit}
\end{equation}
with 
\begin{equation}
x=a_1(r+a_2)^{a_3}
\label{arg}
\end{equation}
This form is chosen because of the asymptotic form of the Bessel functions. 
\begin{equation}
\lim_{x\rightarrow \infty} J_0(z)=\sqrt{\frac{2}{\pi z}}\cos (z-\pi/4)
\end{equation}
and with the constants to be determined in equations (\ref{bfit}) and (\ref{arg}) this form can be made to correspond with the asymptotic behaviour for the current found by Penz {\it et. al.} of $j(x)\propto \sin (x^2)/x$. The results for the eigenvalue are shown in Figure (\ref{fig4}). For this fitting the parameter space appears to have much the same properties as that we found for the Airy function fit. For reference the upper dark blue line is the maximum backflow shown in Figure \ref{fig1}. The green line is the maximum backflow attained with an eigenfunction of the form of equation (\ref{bfit}). At around $\epsilon =0.9$ this is over $99\%$ of the maximum possible value. The values of the parameters that yield this value are $(a_1,a_2,a_3,a_4,a_5,a_6)=(-1.347,0.603, 0.986,0.341, 0.435, 0.715)$. The cyan line is the backflow eigenvalue calculated using the best fit of the form of equation (\ref{bfit}) to the numerically determined eigenfunction. In the relativistic regime above $\epsilon \approx 0.8$ the green and cyan lines coincide to within the level of the numerical accuracy of the calculation. Below $\epsilon \approx 0.8$ it is possible to find a function of the form of equation (\ref{bfit}) that gives greater backflow than the best fit to the numerical data and so the green and cyan lines diverge for low values of $\epsilon$ although, surprisingly they come together at $\epsilon=0.2$. The red line is the same as the cyan line except that we have set $a_6=2/3$. In the region $0.8 < \epsilon < 1.2$ this has essentially no effect on the backflow. For higher values of $\epsilon$ this limitation causes a small decrease in the amount of backflow as one might expect. For low values of $\epsilon$ this line follows the cyan curve very closely, thus showing that for all values of $\epsilon$, but for low $\epsilon$ in particular, the variations in $a_6$ are unimportant. 

\begin{figure}[h]
\centering
\includegraphics[width=0.90\textwidth]{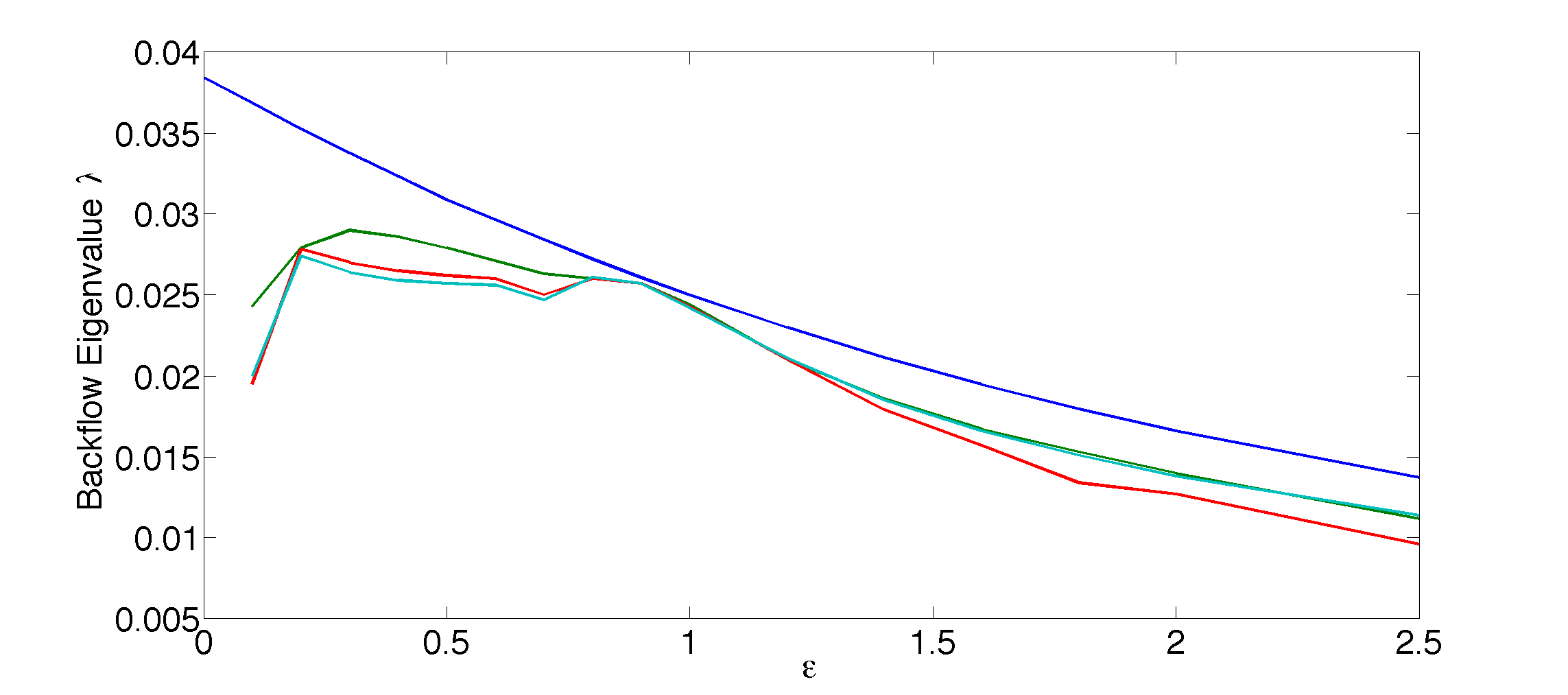}
\caption{{\it Upper line (dark blue)}: the highest magnitude backflow eigenvalue as a function of $\epsilon$ calculated using the exact eigenvector, {\it second line (green)}: The highest magnitude backflow eigenvalue calculated using an optimised value of  equation (\ref{bfit}) as the eigenvector, {\it third line (cyan)} the backflow given by the best fit of the form of equation (\ref{bfit}) to the numerically determined eigenfunction, {\it fourth line (red)} the backflow given by the best fit of the form of equation (\ref{bfit}) to the numerically determined eigenfunction with $a_6=2/3$ }
\label{fig4}
\end{figure}

\begin{figure}[h]
\centering
\includegraphics[width=0.45\textwidth]{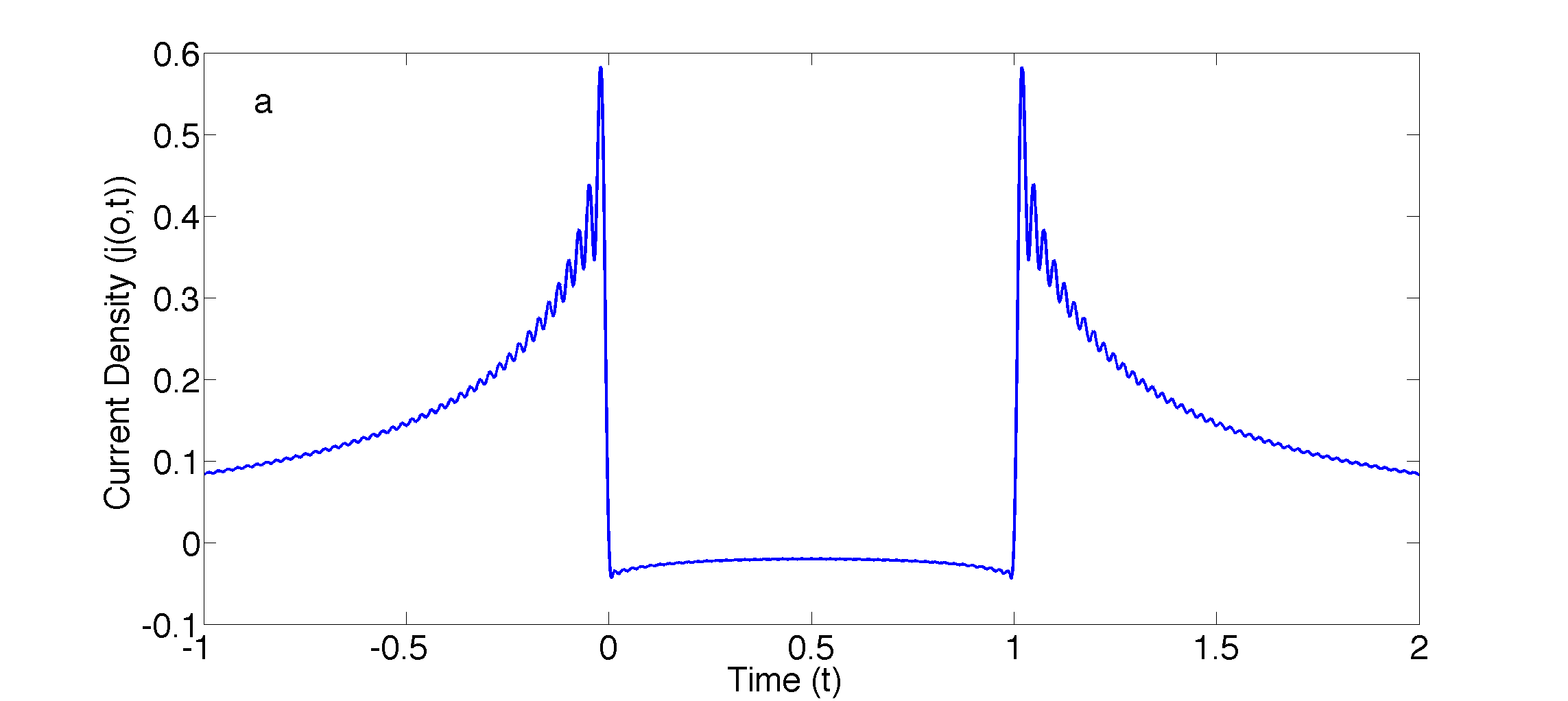}\includegraphics[width=0.45\textwidth]{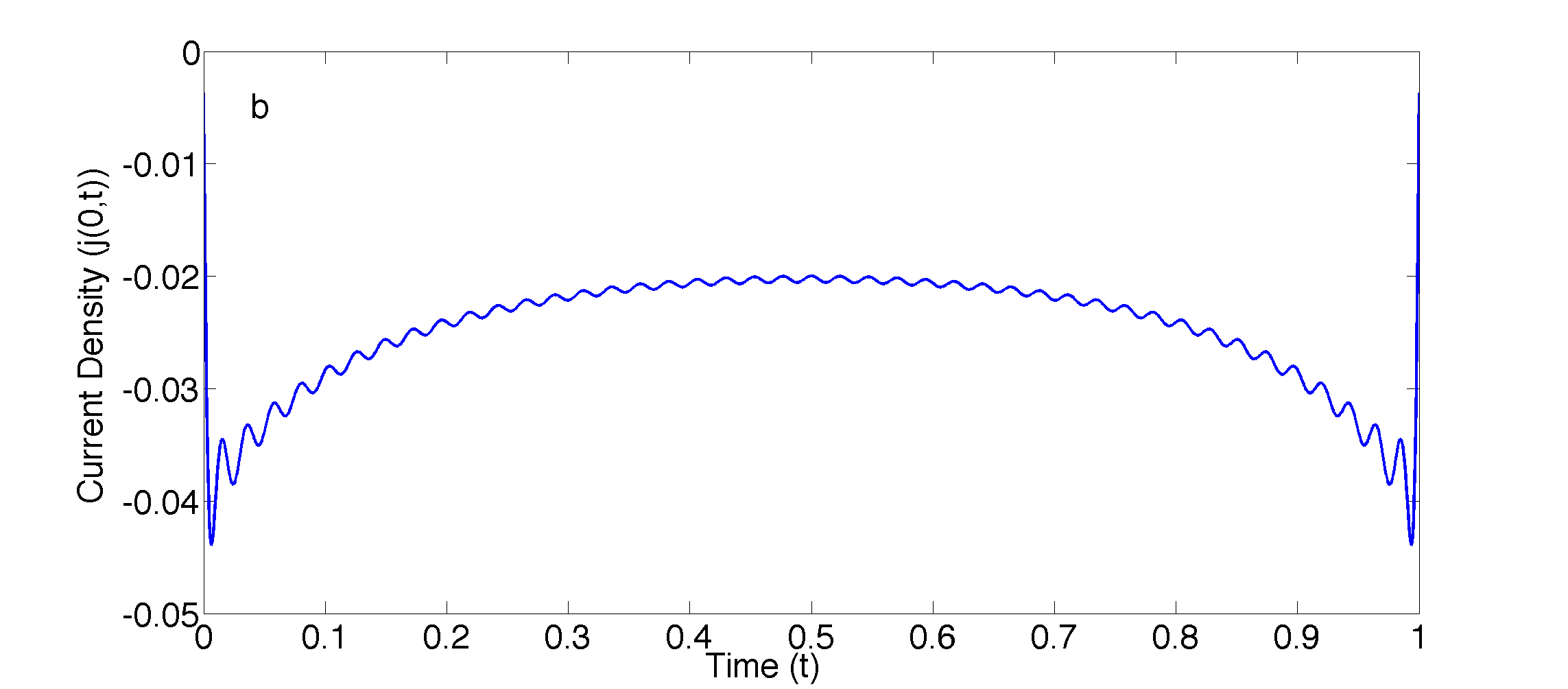}
\includegraphics[width=0.45\textwidth]{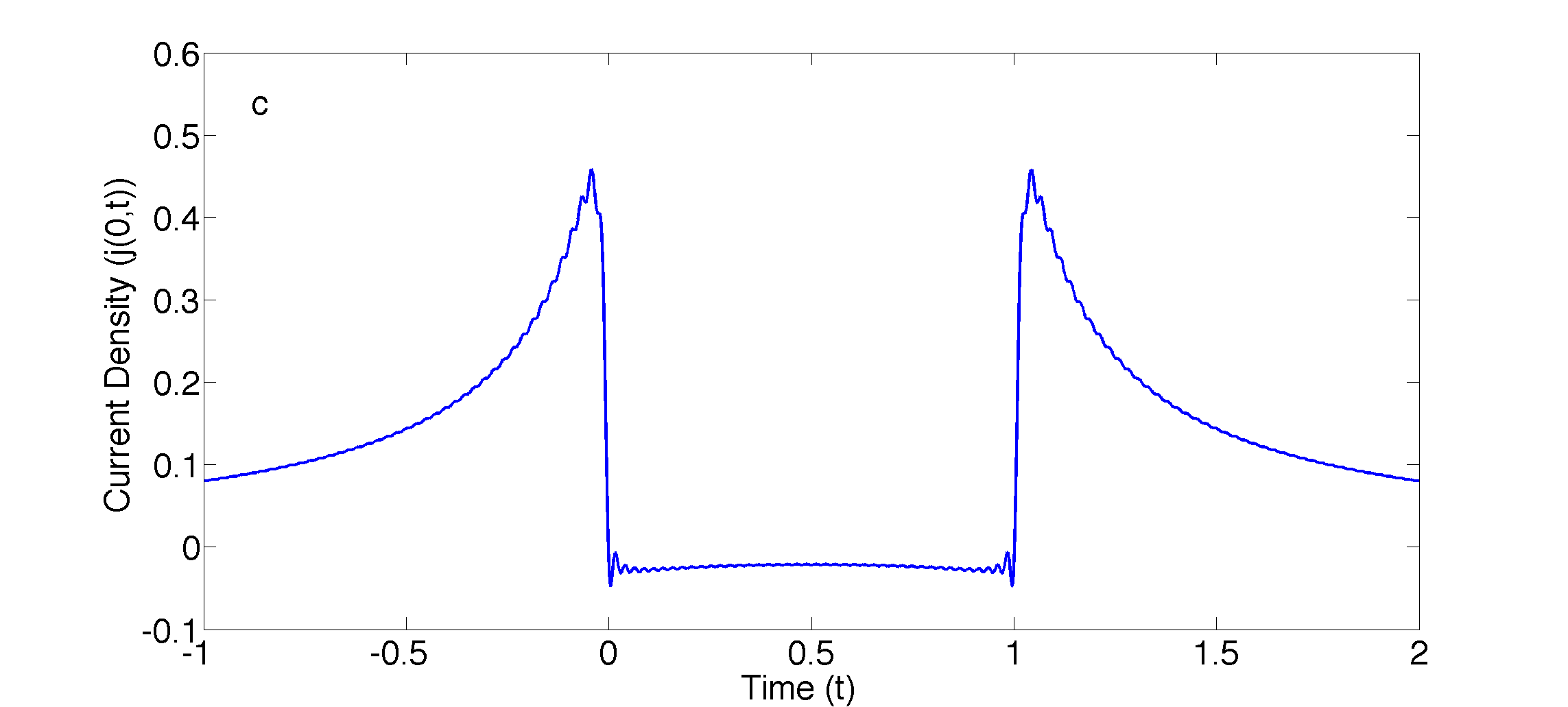}\includegraphics[width=0.45\textwidth]{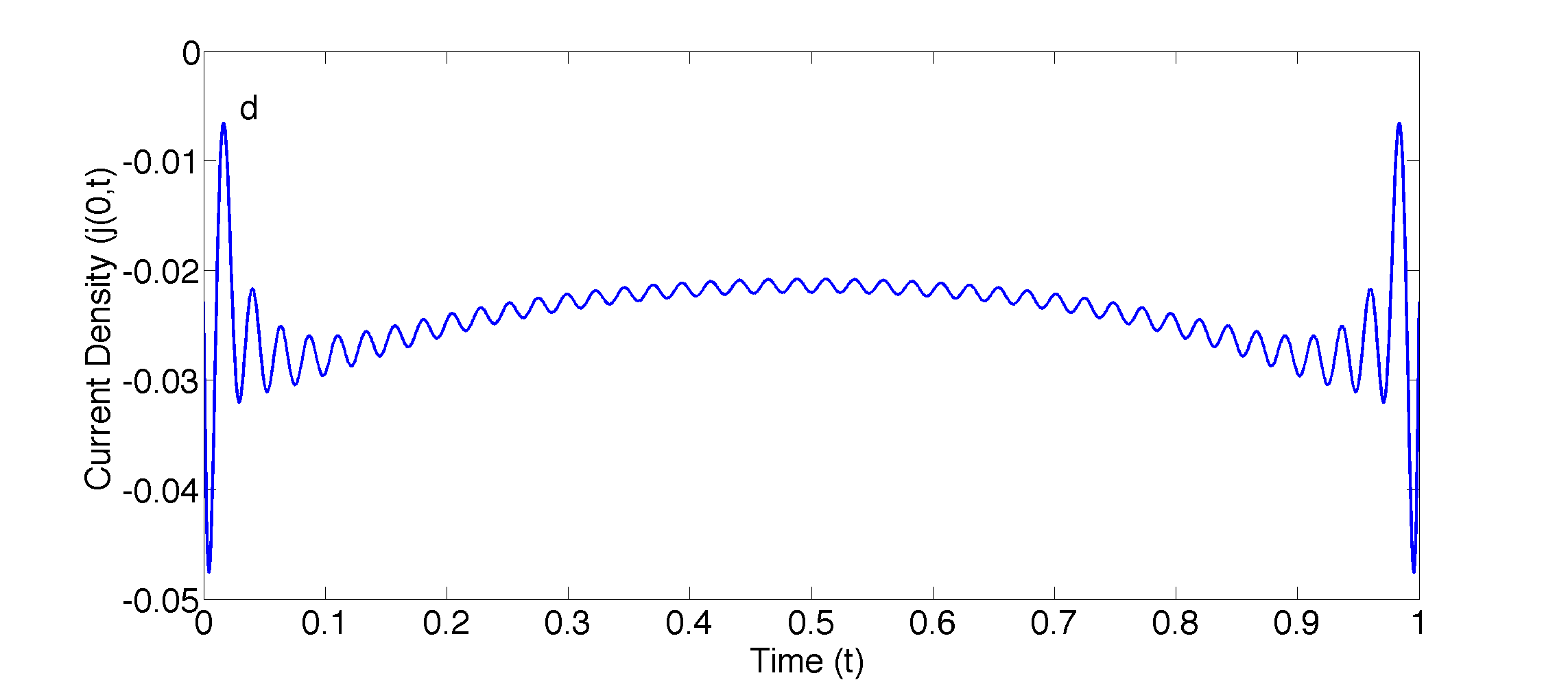}
\caption{{\it a}: The current density as a function of time calculated using equation (\ref{curr}) and the "exact" numerical eigenfunction determined from equation (\ref{eigen1}) for $\epsilon=1.0$, {\it b}: A magnified view of the region of negative current density as a function of time calculated using the "exact" numerical wavefunction, {\it c}: The current density as a function of time calculated using equation (\ref{curr}) and the approximate eigenfunction of equation (\ref{bfit}) with $(a_1,a_2,a_3,a_4,a_5,a_6)=(-1.176, 0.763, 0.971, 0.332, 0.445, 0.751)$, {\it d}: A magnified view of the region of negative current density as a function of time calculated using the approximate eigenfunction.}
\label{fig5}
\end{figure}

The $a$-coefficients are varied to find the function of $r$ which yields the maximum backflow. This is a difficult numerical problem and ultimately we cannot guarantee that we have found the global minimum rather than a local minimum. In fact a number of different sets of coefficients give the same maximum backflow to within the accuracy of the numerical procedures. 

In Figure \ref{fig5} a) and c)we show the current as a function of time for the "exact" numerical eigenfunction and for the optimised function of the form of equation (\ref{bfit}). Figures \ref{fig5} b and d show a magnified view of the current as a function of time in the backflow region. While the upper diagrams resemble to lower diagrams in this figure there is a lot of difference in detail. 

It is clear from Figure \ref{fig3} that the fit of our Airy function and Bessel function wave functions to the exact wavefunction falls off markedly in the non-relativistic limit. In Figure \ref{fig6} we show the current density as a function of time for low and high (non-relativistic and relativistic) values of $\epsilon$ around the region of negative backflow. We controlled $\epsilon$ using $T$ so, although the scales are very different on the horizontal axes in these diagrams, that is of no significance.
\begin{figure}[h]
\centering
\includegraphics[width=0.45\textwidth]{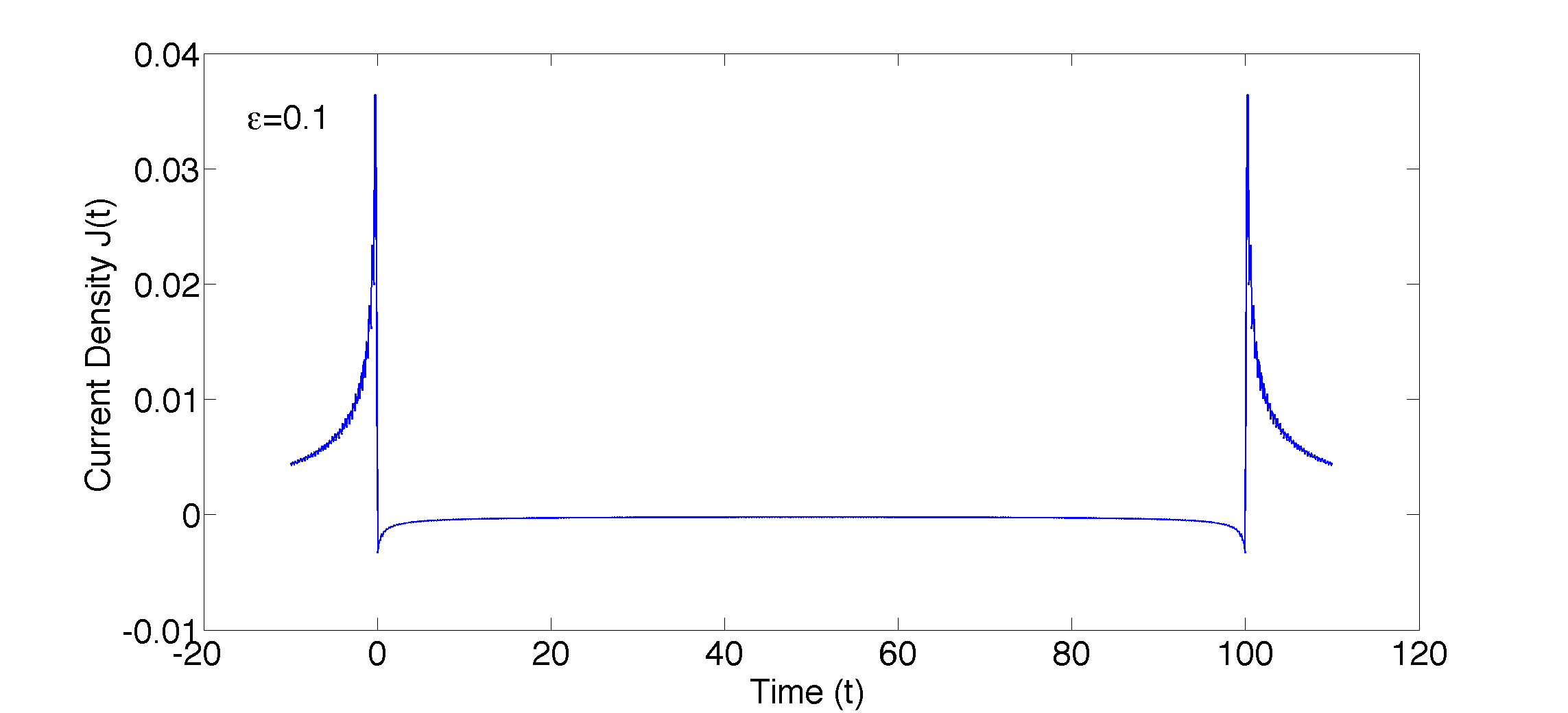}\includegraphics[width=0.45\textwidth]{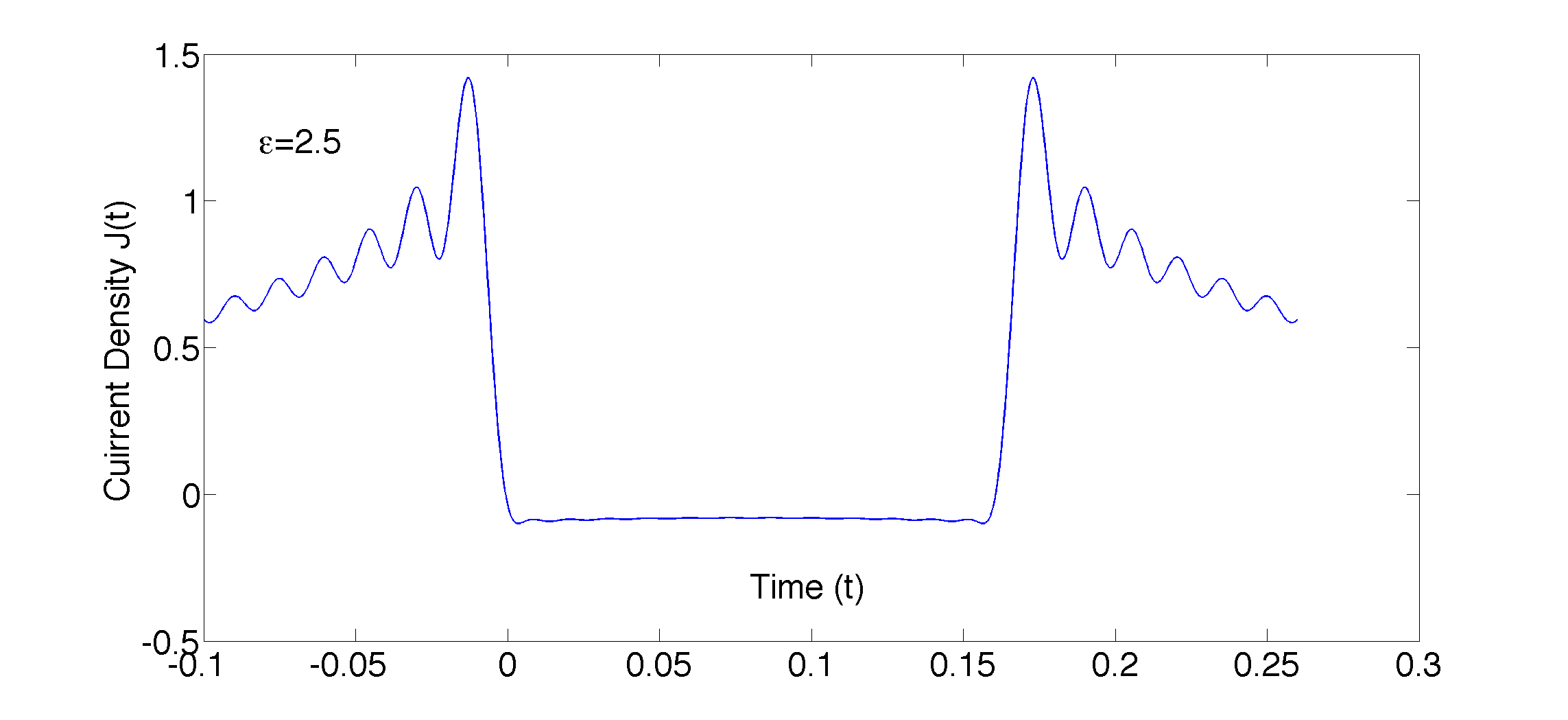}
\caption{The current density as a function of time calculated using equation (\ref{curr}) and the "exact" numerical eigenfunction determined from equation (\ref{eigen1}) for (a) $\epsilon=0.1$; (b) $\epsilon =2.5$}
\label{fig6}
\end{figure}
It is clear that the low $\epsilon$ case is very spiky and the current density is tending towards the singular behaviour observe by O'Mullane and Helliwell \cite{om} in the pure non-relativistic case. On the other hand the current density in the large $\epsilon$ case is relatively smooth and will be much more straightforward to fit with standard functions. In our calculation of the current we have by-passed the co-ordinate space wave functions, but clearly they will be easier to calculate for higher values of $\epsilon$. 

\section{Discussion}

There are a couple of remarkable things about these results. Figures \ref{fig3} and \ref{fig4} show qualitatively the same behaviour for different fitting functions. At large $\epsilon$ the fit using the Bessel functions and the Airy functions yields the same result for the backflow eigenvalue even though the fitting parameters are very different. In fact they are the same to two significant figures for all values of $\epsilon$ we have considered. Both fits are at their best around $\epsilon=0.9$, where they are very good indeed. The rapid fall-off of the curve at low values of $\epsilon$ is consistent with earlier work which found that attempting to fit the non-relativistic theory of backflow with a Bessel function or an Airy function resulted in only modest backflow. 
There is a minimum in the backflow eigenvalue at $\epsilon=0.7$  for both the Airy function and the Bessel function fits to the numerical data. We were unable to find a higher value for the backflow eigenvalue at $\epsilon=0.7$ for these trial functions despite repeated searching. Furthermore the various levels of approximation come together and give identical backflow in a small region around $\epsilon=0.2$. This also happens for both the Airy and Bessel fitting functions. We are forced to leave these as unexplained observations, but are currently investigating them further. For both cases finding the maximum backflow at low values of $\epsilon$ was much more difficult than for high values because a number of local minima of roughly equal depth appear in the parameter space in this limit. It is important to get the fitting right at low values of $r$, $r=1$ corresponds to a momentum $p=mc$, and high values of $r$ represent the tail of the distribution of momenta. The bulk of the weight of the momentum space wavepacket is in the region $r<1$. The backflow in the trial function is a maximum around $\epsilon=0.9$ because that is the region in which the trial function gives the closest to perfect fit for momenta below $r \approx 1$. 

\section{Conclusions}

It has been shown in equation (\ref{fit}) that the maximum relativistic backflow eigenvalue can be written in terms of the non-relativistic backflow eigenvalue and the parameter $\epsilon$ over a wide range of values of $\epsilon$. We have also seen that a fit to the backflow eigenfunction in terms of Bessel functions of the first kind gives over $99\%$ of the maximum possible backflow at optimal values of $\epsilon$. From figures \ref{fig3} and \ref{fig4} we can see that in the non-relativistic limit $\epsilon \rightarrow 0$ the fitting of these functions to the numerical data becomes very poor and results in the fitted functions yielding a much lower backflow than the theoretical maximum. Furthermore, as we tend to the non-relativistic limit, a number of minima in the $a_j$ parameter space of approximately equal depth emerge. This has been observed previously in the  non-relativistic theory, but here we see that it does not extend into the relativistic regime. This means that this is one of those rare problems where the relativistic theory is intrinsically more easy to solve than the non-relativistic theory. This may well be of relevance for attempts to observe backflow experimentally. Finally we point out that here we have considered spatial backflow. In relativistic theory, space and time should be treated on an equal footing and so temporal backflow should also exist. This has been investigated and is interpreted in terms of the negative energy solutions of the Dirac equation by Su and Chen \cite{such}. 

\section{Acknowledgements}
We would like to thank the University of Kent for their support. We would also like to thank Professor M. Penz for allowing us access to his backflow software.

\end{document}